\documentclass[12pt,preprint]{aastex}

\usepackage{graphicx}

\shorttitle{On the Physical Problem of Spatial Dimensions: An Alternative Procedure to Stability
Arguments}
\shortauthors{F. Caruso \& R. Moreira}

\begin{document}

\title{On the Physical Problem of Spatial Dimensions: \\
An Alternative Procedure to Stability Arguments }

\author{Francisco Caruso \& Roberto Moreira Xavier}
\affil{Centro Brasileiro de Pesquisas F\'{\i}sicas \\
Rua Dr. Xavier Sigaud 150, 22290-180, Rio de Janeiro, RJ, Brazil}
%\email{caruso@cbpf.br}

\email{caruso@cbpf.br}


\begin{abstract}
Why is space 3-dimensional? The first answer to this question, entirely
based on Physics, was given by Ehrenfest, in 1917, who showed that the
stability requirement for $n$-dimensional two-body planetary system very
strongly constrains space dimensionality, favoring 3-d. This kind of
approach will be generically called ``stability postulate" throughout this
paper and was shown by Tangherlini, in 1963, to be still valid in the
framework of general relativity as well as for quantum mechanical hydrogen
atom, giving the same constraint for space-dimensionality. In the present
work, before criticizing this methodology, a brief discussion has been
introduced, aimed at stressing and clarifying some general physical aspects
of the problem of how to determine the number of space dimensions. Then,
the epistemological consequences of Ehrenfest's methodology are critically
reviewed. An alternative procedure to get at the proper number of
dimensions, in which the stability postulate (and the implicit
singularities in three-dimensional physics) are not an essential part of
the argument, is proposed. In this way, the main epistemological problems
contained in Ehrenfest's original idea are avoided. The alternative
methodology proposed in this paper is realized by obtaining and discussing
the $n$-dimensional quantum theory as expressed in Planck's law, de
Broglie relation and the Heisenberg uncertainty relation. As a consequence,
it is possible to propose an experiment, based on thermal neutron
diffraction by crystals, to directly measure space dimensionality. Finally
the distinguished role of Maxwell's electromagnetic theory in the
determination of space dimensionality is stressed.
\end{abstract}

\section{Introduction}\label{int}

          In this paper we discuss the dimensionality of space as a
physical problem.

          At first sight this problem could be approached, in a fruitful
way by simply asking the question: why is space three-dimensional? However,
on second thought, it is clear that this formulation is narrow minded since
the three-dimensionality of space is assumed as something given {\it a
priori} perhaps by our sense organs, especially vision. We shall come back
to this point later, but now it suffices to say that our interaction with
the external world (via our vision) is essentially electromagnetic and that
Electromagnetism implies a three-dimensional world, as we will see below.
Now, in Physics we are not restricted to our direct experience of the
external world, {\it i.e.}, to our sensory perception. So we can
investigate the problem in a much more profound way by freeing ourselves
from our sensory prejudices and trying to answer the following
complementary questions: {\it i)} How does it become manifest in the
fundamental laws of Physics that space {\it has} 3 dimensions? {\it ii)}
How do the fundamental laws of Physics {\it entail} space dimensionality?
These two questions will be discussed throughout this paper both by
critically reviewing the existing literature and by proposing new
approaches to this problem.

          Some readers may find somewhat ``unpleasant" in this paper the
several digressions and some apparently ``unnecessary" repetitions. We hope
that this feeling will disappear by the time we come to the conclusions,
with the realization that these digressions and footnotes are indeed
necessary and that, often, they are there to clarify points which would,
otherwise, remain somewhat obscure. In this perspective, they are, in fact,
fundamental to our final discussion.

          Before proceeding and entering more deeply into our subject we
must first clarify some points which more plainly define our conceptual
framework.
  We begin by considering that the dimensionality of space {\it is
not a contingent feature}. To accept this means that one must search for a
general methodology capable of determining it. A fundamental ingredient is
necessarily the possibility of thinking about higher dimensional space,
which is provided by the works of Lobacevskij, Bolyai, Gauss, Cayley,
Grassmann, and Riemann [\cite{Jammer}], but as we will see below this is not
sufficient. Although at early times the physical soundness of this kind of
generalizations was continuously questioned,\footnote{An example of criticism where the three-dimensionality of space is
considered as a contingent feature can be found, for example, in Mach E.,
{\it Die Mechanik in ihrer Entwicklung historisch-kritisch dargestellt},
Leipzig, 1883, Italian Transl., by A. D'Elia, {\it La Meccanica nel suo
sviluppo storico-critico}, Torino, Boringhieri, 1977, pp.~479-80.} there is
nowadays a kind of general consensus that theories in higher dimensions
({\it when supplemented} with dimensional reduction) may provide a
promissing framework for a deeper understanding of very high-energy
physics. However, it is clear that the very fact of {\it imposing}  the
process of dimensional reduction in a given higher dimensional theory is
equivalent to assuming {\it a priori} the dimension number 3 as a {\it
natural} property of space, which is just what we are querying. To the best
of our knowledge, there is as yet no satisfactory and unambiguous answer to
the problem of dimensional reduction in the framework of these theories,
even when the so called spontaneous compactification process is taken into
account.\footnote{It is shown that spontaneous dimensional reduction
in any Kaluza-Klein theory always yields a compactified extra space.
However, without and adjustable cosmological constant, the scale of the
ordinary four-dimensional space-time is the same order of magnitude as
that of the compactified space.  {\it Cf.} Tosa, Y., ``Spontaneous
dimensional reduction in Kaluza-Klein theories", {\it Phys. Rev.} {\bf
D30} (1984) 339; See also Cremmer, E. and Scherk, J., ``Spontaneous
compactification of space in an Einstein-Yang-Mills-Higgs model", {\it
Nucl. Phys.} {\bf B108} (1976) 409. Now for illustrating the present
difficulties on this subject, concerning superstring theory, we can quote
Ferrara's words: ``{\it Superstring are 10-dimensional theories of
one-dimensional extended objects, so their relation to the physical world
is only possible if they undergo a mechanism of spontaneous
compactification fron $D=10$ to $D=4$ dimensions. The study of spontaneous
compactification of the fully fledged superstring theory is a formidable
task to achieve, since it requires the knowledge of the full
second-quantized version of the interacting theory.}" {\it Cf.} Ferrara, S.
``Matter Coupling in Supergravity", {\it in} {\it Superstring and
Supergravity -- Proceedings of the Twenty-Eighth Scottish Univ. Summer
School in Physics}, A.T. Davies and D.G. Sutherland (eds.), Oxford, Univ.
Printing House, 1985, p.~381.} Thus we need to propose some physical argument to introduce
another fundamental ingredient which, together with the former, will allow
us to {\it start} the discussion of whether this number is indeed 3 -- but
not necessarily to {\it determine} it. This ingredient will be provided by
the realization that a particular physical law is intimately dependent on
the number of space dimensions. Historically, Kant's conjecture [\cite{Handyside}] that
the three-dimensionality of space may, in some way, be related to Newton's
inverse square law has, indeed, opened a new way for the study of the
problem of space dimensions. The main contribution of this conjecture to
this problem is thus the suggestion that it can also be treated as a {\it
physical} problem and does not belong exclusively to the domain of
mathematics. It is relevant to stress here that, in spite of the importance
of this conjecture, its physical support (if any) is yet to be understood.

          Usually a third (and decisive) ingredient is always required to
suggest a method which effectively connects the number of dimensions to
some physical property. This is the most delicate part of any method one
can propose for discussing the problem of spatial dimensions, which will be
carefully examined throughout this paper. Here, only the physical aspects of
this problem are discussed and, in particular, epistemological consequences
of Ehrenfest's methodology aimed at fixing the number of space dimensions
base on the so called ``stability postulate" (see Section~\ref{learn}) are critically
discussed. Some of the fundamental ideas related to the physical nature of
this problem and to the question of the physical relevance of spatial
dimension -- treated from different points of view [\cite{Poincare}], [\cite{Buchel}], [\cite{Grunbaum}], [\cite{Barrow}] -- will also be
briefly reviewed in Section~\ref{learn} but, before discussing any principle that
could be used to determine space dimensionality, we would like to say that
we are convinced that it is impossible to disentangle questions concerning
this subject from some kind of formalism representing a physical law. As
Jammer put it [\cite{Jammer2}],

\begin{quotation}
``{\it ... Hence it is clear that the structure of the
space of physics is not, (...), anything given in nature or independent of
human thought. It is a function of our conceptual scheme.}"
\end{quotation}

This means that
we accept that the physical concepts and the concept of reality itself
acquire sense only within a theoretical construction where they can be
discussed and realized. When the problem of space dimensions is considered,
we must carefully examine the consequences of this fundamental point.
Although this point has, in fact, motivated several works on the problem of
spatial dimensions, it is in itself, at the same time, one of the main
difficulties for the discussion of the problem, because the
three-dimensionality of space is not questioned {\it a priori} when a
physical law is established. This essential difficulty would be bypassed if
we are able to prove the validity of the physical law in question whatever
the number of spatial dimensions under consideration, rather than simply
postulating it. The main aim of this paper is exactly to develop this point.

          Concerning the origin of the results one may arrive at by
discussing the {\it problem of the number of dimensions} in the way
prescribed above, there is a straightforward and very important consequence
we would like to emphasize, namely: {\it the mathematical structure of the
formalism} one is considering (or simply a given physical equations) is
the {\it causa formalis} of the constraint obtained on the number of space
dimensions. Actually we tend to consider this as the {\it unique} approach
to start discussing the problem of space dimensionality and this is
essentially related to Jammer's idea recalled above. Thus this
epistemological limitation seems to be inherent to this problem (so far as
we understand it) and, in a certain sense, is well illustrated by
Grassmann's words [\cite{Grassmann}]:

\begin{quotation}
{\noindent \it ``The concept of space can in no way be produced by thought,
but always stands over against it as a given thing. He who tries to
maintain the opposite must undertake the task of deducing the necessity of
the three dimensions of space from the pure laws of thought, a task whose
solution presents itself as impossible."}
\end{quotation}

          This paper is organized as follows. In Section~\ref{learn} the present
status of what we can learn from the formal extension of the number of
space dimensions is discussed. Particular attention is given to Ehrenfest's
and Weyl's contributions to this subject. A brief comment on the reality
criterion associated with the  ``extra dimensions" in theories at higher
dimensions is also presented in this Section~\ref{learn}. As a result of the
criticism of the use of the ``stability postulate", carried out in Section~\ref{criticism}, an alternative approach to get at the proper dimensionality of space is
presented in this same Section. In Section~\ref{black} it is shown how the task
proposed in Section~\ref{criticism} can be carried out by considering a particular
transition $\Re^3 \rightarrow  \Re^n$ for the case of the black body
phenomenology. This enable us to ``demonstrate" the validity of the de
Broglie relation for any $\Re^n$. This is the basis of Section~\ref{thermal}, where
thermal neutrons diffraction by crystals is presented as an example that
completes the procedure proposed in Section~\ref{criticism}. An upper limit for the
dimensionality of space is therefore obtained. Some conclusions are drawn
in Section~\ref{conclusion}.

\section{What one expects to learn from the
transition $\Re^3 \rightarrow  \Re^n$}\label{learn}

          As a first example, we can quote Ehrenfest's fundamental papers [\cite{Ehrenfest17}], [\cite{Ehrenfest20}]. There, several physical phenomena, where qualitative differences
between three-dimensional $(\Re^3)$ and other $n$-dimensional $(\Re^n)$
spaces were found, have been discussed. These aspects, which distinguish
the $\Re^3$ Physics from the $\Re^n$ one, are called by him ``singular
aspects" and his works were aimed at stressing them. A crucial assumption
is built in the main ideas contained in [\cite{Ehrenfest17}], [\cite{Ehrenfest20}], namely, that it is possible to
make the formal extension $\Re^3 \rightarrow  \Re^n$ for a certain law of
Physics and, then, find one or more principles which, in conjunction with
this law, can be used to single out the proper dimensionality of space. For
this approach to be carried out, in general, the form of a differential
equation -- which usually describes a physical phenomenon in a
three-dimensional space -- is maintained and its validity for an arbitrary
number of dimensions is {\it postulated}. For example, the Newtonian
gravitational potential for a $\Re^n$-space, $V(r) \propto r^{2-n}$, is
the solution of the Laplace-Poisson equation,

$$ \sum\limits^{n}_{i=1} = {{\partial^2 V}\over {\partial x_i^2} }= k \rho,$$

\noindent in an $n$-dimensional space. Based on this general solution,
Ehrenfest has used the postulate of the stability of orbital motion under
central forces to get at the proper number of dimensions. In Ehrenfest's
approach this additional postulate acts, therefore, as the {\it causa
efficiens} of the dimensionality of space. It is just this part of his
method the object of the criticism in Section 3.

          This general procedure is also followed in the work of Whithrow
[1955, 1959]. The importance of this approach was noted by Tangherlini [1963, 1986] who
proposed that, for the Newton-Kepler (N.K.) problem generalized to $\Re^n$
space the principle to determine the spatial dimensionality could be
summarized in the postulate that there should be stable bound states orbits
or ``states" for the equation of motion governing the interaction of
bodies, treated as material points. This will be generically called from
now on, the {\it stability postulate}. In his first paper [1963], Tangherlini
showed that the essential results of the Ehrenfest-Whitrow investigation
are unchanged when Newton's gravitational theory is replaced by general
relativity. In this same paper, the Schr\"odinger equation for the hydrogen
atom in $n$ dimensions is also considered. The above postulate, in
conjunction with the assumption that the fields produced by the nucleus
asymptotically approach a constant value at ``large distances", gives $n=3$
in both cases. Thus the three-dimensionality of space discussed within the
framework of Newtonian mechanics [\cite{Ehrenfest17}], [\cite{Ehrenfest20}], [\cite{Whitrow}] or general relativity [\cite{Tangherlini63}], and also
quantum mechanics [\cite{Tangherlini63}] (using a Coulombian potential), seems to be a
result valid for a very large range of spatial scale -- we will return
to this point in Section~\ref{conclusion}. This briefly reviews how the ``stability
postulate" is used to throw some light on the problem of spatial
dimensions.

          From another point of view, these attempts based on stability
arguments belong to a class of arguments epistemologically different from
that contained in the work of Weyl [1918, 1919, 1952], which we shall briefly review here.
His basic approach was to construct a new unified theory of gravitation and
electromagnetism based on a gauge-invariant non-Riemannian geometry. In
this scheme, Weyl pointed out that there is a strong relation between the
metric structure of space-time and physical phenomena, which could lead to
a deeper understanding of Maxwell's electromagnetic theory as well as of
the four-dimensionality of space-time. Weyl showed that only in a
(3+1)-dimensional space-times can Maxwell's theory be derived from a
simple gauge-invariant integral form of the action, having a Lagrangean
density which is conformally invariant. This could be considered as an
example of how a set of physical phenomena, here synthesized by Maxwell's
theory, could be used to impose some restrictions on the dimensionality of
space.\footnote{Indeed, this result is based on classical arguments and one can argue
that this is not the only example. In fact, one gets the some constraint on
$n$ when extending Weyl's approach to classical Yang-Mills theory -- Yang,
C.N and Mill, R.L., ``Conservation of Isotopic Spin and Isospin Gauge
Invariance", {\it Phys. Rev.} {\bf 96} (1954) 191.} The structure of Maxwell equations and the gauge principle are,
respectively, the {\it causa formalis} and the {\it causa efficiens} of the
four-dimensionality of space-time. The two essentially different
(although complementary) features of Ehrenfest's and Weyl's methodology can
be summarized as the difference between the two following questions: (i)
How does it become manifest in the fundamental laws of Physics that space
has three-dimensions, and (ii) How do the fundamental laws of Physics
entail spatial dimensionality? All work based on the ``stability postulate"
hinges on the former question because the constraint on $n$ is reached as a
consequence of a ``singular aspect" of a physical law that is {\it supposed
to be still valid} under the transition $\Re^3 \rightarrow \Re^n$. The
latter is implicit in Weyl's work where the structure of Maxwell theory
{\it cannot} be maintained\footnote{{\it What inner peculiarities distinguish the case $n=3$ among all
others?  If God, in creating the world, chose to make space 3-dimensional,
can a reasonable explanation of this fact be given by disclosing such
peculiarities?}, {\it cf.} Weyl, H. in {\it Philosophy of Mathematics and
Natural Science}, revised and augmented English transl., by O.~Helmer,
Princeton, Princeton Univ. Press, 1949, p.~70. Weyl has shown that
electromagnetism plays such a particular r\^ole; {\it cf.} {\it ibid.} pp.~136-37 and ref.~[\cite{Weyl}].} if $n \not= 3$. The second question can be
well illustrated by the concluding paragraph of Tangherlini's paper [1963],
where he says: {\it with further work, we may come to regard $n=3$ as an
eigenvalue}.

          However, even from a classical point of view, Weyl's
demonstration of the four dimensionality of space-time is not complete:
the gravitational law should also be derived from the same requirements of
invariance as for electromagnetism. The point is that although Weyl's
unified theory is a good place for giving ans answer to the problem of
spatial dimensions, it should be mentioned that this theory has been
criticized in the literature [\cite{Bergmann}], [\cite{Dirac}]. In any case, in order to consider
complete such kind of demonstration, today we must clearly take into
account also {\it strong} and {\it weak} interactions. We will turn later
to this point at the end of this Section.

          Other attempts to create a geometry in which the gravitational
and electromagnetic potentials together would determine the structure of
space were carried out. An example is Kaluza-Klein theory [\cite{Kaluza}], [\cite{Klein}] -- which is
presently enjoying a great revival of popularity in connection with the
modern theories of supergravity -- where the number of components of the
metric tensor was increased by changing the number of spatial dimensions. A
fifth dimension was added to the usual four dimensions of physical
space-time. In the work of Kaluza, the {\it a priori} four-dimensional
character of the physical world is assumed when the author looks for a
suitable choice of coordinates, in such a way that the components of the
metric tensor be independent of the fifth coordinate. In other words, this
coordinate has no direct physical significance. Thus it is quite clear that
this kind of approach to a unification program could not lead to a
satisfactory answer to the problem of spatial dimensions.\footnote{See footnote 2.} However it
should be said that an argument aimed at showing that a necessary condition
for a unified field-theoretic description of gravity and electromagnetism
implies that the world be four-dimensional, as discussed by Penney [1965].
The four dimensionality of space-time is also required by Sch\"onberg [1971]. In this interesting work and electromagnetic foundation for the
geometry of the world-manifold is proposed. Einstein's gravitational
equation appears as complementing the set of Maxwell equations, giving rise
to a natural fusion of the electromagnetic and gravitational theory. The
electromagnetic theory is formulated in a differentiable manifold devoid
of any metric and affine structure. In this formulation there is no {\it a
priori} relation between $F_{\mu\nu}$ and $F_{\mu\nu}^\ast$, involved in
the homogeneous and non-homogeneous Maxwell equations, respectively. The
foundation of the four-dimensionality of the world-manifold (space-time)
is given by the structure of the Maxwell equations in terms of two basic
tensors $F_{\mu\nu}$ and $F_{\mu\nu}^\ast$, which are both antisymmetric
covariant of the same order. It is important to stress that, in this
approach, the four-dimensionality of the space-time is essentially
associated to the differential electromagnetic equations, without any
consideration about the relation between $F_{\mu\nu}$ and $F_{\mu\nu}^\ast$
and without requiring a metric-space as in Weyl's work.

          There are other attempts to unify not only electromagnetic and
gravitational forces but all the fundamental forces, considering
space-time with a high number of dimensions as, for example, supergravity
or the ten-dimensional space-time superstring theory [\cite{Niewenhuysen}], [\cite{Schwarz}]. But, whenever
the problem of space-time dimensionality is considered in the framework of
these (super-) theories, we face the problem of the physical reality of
these ``extra" dimensions. Independently of any particular theory, as
pointed out by Mansouri and Witten [1985],
\begin{quotation}
{\noindent \it ``if we wish to take the physical existence of the extra
dimensions seriously, we must develop a systematic method for studying the
effects of the extra dimensions (...) Since there is no evidence for the
existence of the extra dimensions at the shortest distance which can be
probed at present, it [any such method] must explain how this can be
attributed to some intrinsic property of a higher dimensional theory. It
must [also] provide a quantitative method for studying the consequences of
the dependence on the extra dimensions."}
\end{quotation}

          Complementing this picture we can always ask whether the
ten-dimensional superstring theory, for example, can tell us, in a
straightforward and unambiguous way, that we are living in a (``almost
flat") four dimensional space-time. The fundamental question is: {\it why}
dimensional reduction? Up to now, the answer to this question, {\it i.e.},
the four-dimensionality of the physical world-manifold, is yet, in the
last analysis, an {\it ad hoc} ingredient in these theories.

          On the other hand, it was shown [\cite{Barrow}] that only for space-time
dimensionality greater than four, the fundamental constants of
electromagnetism ($e$), quantum theory ($\hbar$), gravity ($G$) and
relativity ($c$) are all included in a single dimensionless constant --
which should have, in a unified theory, a similar r\^ole to that played by
the Sommerfeld constant $e^2/\hbar c$ in the quantum electrodynamic theory.
Thus the apparent necessity of going to a high dimensional space-time, in
order to carry out the unification program, brings with itself the problem
of how to explain all the well-known phenomenological manifestations of the
four-dimensionality of space-time in the framework of this new theory,
and the question of the reality of the ``extra-dimensions": both are
clearly still open questions in Physics.

\section{Criticism of the use of stability
postulate}\label{criticism}

          We can ask if the ``stability postulate" -- applied to the N.K.
problem or hydrogen atom -- is actually a good choice for deriving the
spatial dimensionality or not; or more specifically, if we can really {\it
prove} that $n=3$. We understand that the use of this postulate enables us
only to exclude the possibility of having a class of natural phenomena in a
space other than our own, with an arbitrary dimension, as pointed out by
Poincar\'e [\cite{Poincare17}]. Then, when we consider the example of the hydrogen atom,
as described in Section~\ref{learn}, the results obtained from that postulate must be
stated as follows: there is no $\Re^n$ {\it other than} $\Re^3$ where the
phenomenon under study is described by a generalized Schr\"odinger equation
that has the same form as in the case $n=3$, and whose solution is {\it
also} stable -- and that is all. Indeed, when Ehrenfest used the Bohr
atomic model for the hydrogen atom, the stability of matter in three
dimensions was {\it already} assured by the postulate of angular momentum
quantization, and this justifies the term {\it also} underlined above. The
fact is that he could not have used Rutherford's model -- which is clearly
unstable in $\Re^3$ -- plus the stability postulate to derive the number of
dimensions as being just 3. Thus $n=3$ is {\it a priori} favored in this
case. Apart this feature, it is clear that it is only when the formalism,
previously generalized to an $n$-dimensional space, presents a singular
behavior under this generalization, that the ``stability postulate" can be
used as a method to fix the proper dimensionality of space. The range of
applicability of the ``stability postulate" is therefore strongly
restricted to a very particular class of formalisms. Moreover, these two
intrinsic characteristics of this method clearly do not solve the essential
difficulty discussed in Section 1 and, from the epistemological point of
view, show that the use of the ``stability postulate" to fix $n$ is not
satisfactory.

          We can now ask if we cannot imagine a phenomenon or a physical
state that could only be stable in an $\Re^n$ with $n > 3$, but described
by an equation having the same form as in $\Re^3$, and analyze the
consequences of this assumption. For example, we can ask why we do not
observe in a Stern-Gerlach [1922] experiment the dissociation of a beam od
spin 1/2 particles in more than two lines. Or, in other words, is the
stability of these particles ({\it e.g.} electrons), described by a Dirac
equation, a manifestation of a particular space dimensionality? Particles
having higher spin must be unstable in $\Re^3$, while stable in some $\Re^n
\not= \Re^3$ and so, having a mean lifetime so small in three dimensions,
this kind of experiment could not be carried out. This conjecture could
indicate that if the ``stability postulate" were applied to the evolution
of a massive spin 3/2 particle, described by a (hypothetical) Dirac-like
equation, the number of spatial dimensions derived could be greater than 3!
This is an example where the results obtained by using the ``stability
postulate" do not depend on the form of the equation but, instead, on what
kind of object this equation describes.

          The alternative principle we want to propose may be stated as
follows: ``Given a formalism in a certain dimension, (usually three) we
must, based upon its fundamental equations, ask whether other forms (or
equations) are valid in a higher $n$-dimensional space for {\it all} $n$,
rather than simply postulating the validity of the same formalism in a
different dimension". In other words we shall not be concerned only with
formalisms which are singular in a certain $n$ (usually three). On the
contrary, we shall look for situations which do not present those
singularities. Then this alternative principle could be used to discuss the
spatial dimensionality (Section~\ref{black}). It will certainly describe several
phenomena and their observability could not be used for that purpose. It is
clear, however, that in this case, the constraints obtained will be weaker
than those obtained when the ``stability postulate" or the search for
singular aspects of the transition $\Re^3 \rightarrow \Re^n$ are
considered. Nevertheless, this procedure has the advantage that we can
guarantee {\it a priori} that the fundamental law, used to describe a
certain kind of phenomenon, is valid for any $\Re^n$, which is not possible
in other procedures as pointed out in Section~\ref{int}. Then, when this
alternative procedure is applied we can conclude: the dimensionality of
space {\it is} a number included in a certain range -- 3 need not {\it a
priori} be favored.

\section{Black body phenomenology: a non singular
aspect of the transition $\Re^3 \rightarrow \Re^n$}\label{black}

          There are several physical laws in which the dimensionality of
space affects the results, but the transition $\Re^3 \rightarrow \Re^n$
does not have a ``singular" behavior, and thus these laws were not
discussed in the works of Ehrenfest [1917, 1920]. An example is Wien's law which, in
its generalized form,\footnote{ It should be noted that this generalization follows purely from the
validity of thermodynamics in $\Re^n$, leaving the explicit form of
$F(\nu/T)$ open. See also footnote 8.} becomes $\rho = \nu^n F(\nu/T)$. However, we
would like to point out that, although this transition has no
``singularity", the black body phenomenology extended to $\Re^n$ contains an
important feature that must be emphasized. Indeed we can use it in order to
``demonstrate" the validity of the de Broglie relation in other $\Re^n$, as
will be shown now.

          If we assume Planck's energy quantization to determine the
explicit form of the function $F$, we still find that the energy of a
quantum is $\epsilon_0 = h \nu$, for any $\Re^n$. This is easily seen if we
remember that the energy eigenvalue of the Schr\"odinger equation for the
harmonic oscillator gives Planck's result up to the ground state energy.
The transition to $\Re^n$ only changes this energy value from $3h\nu/2$ to
$nh\nu/2$, and then Planck's hypothesis is still valid, {\it i.e.}, the
quantum energy is proportional to the {\it first} power of the frequency
$\nu$. We note that this result clearly depends on the classical potential
energy $V=kx^2/2$ used in the Schr\"odinger equation, and a brief
digression about it is necessary.

          When a spring is displaced for the equilibrium position, we learn
from the experiment that, for small displacements, the restoring force is
proportional to the displacement, and that is all. It does not matter in
which direction the displacement takes place and the problem can be called
a quasi one-dimensional problem. The result is the same in one, two or
three dimensions and this is quite different from the Newtonian-Keplerian
potential, for which a qualitative difference among $\Re^3$ and $\Re^n$
exists [\cite{Ehrenfest17}], [\cite{Ehrenfest20}]. Thus we can expect from induction that the form of Hooke's
potential could be the same for all $\Re^n$. However, even if this is not
true, but if the generalized potential has a minimum, we can always
approximate it by the harmonic potential, in the case of small
oscillations, whatever $\Re^n$ is considered (a particular case of Morse
theorem). After this note, we can turn back to the original problem.

          We can still assume that the energy trapped in a cavity (a model
for a black body) corresponds to the energy of a collection of ``photons"
which must satisfy Einstein's relation $M^2=g_{\mu\nu} p^\mu p^\nu$,
generalized to $\Re^n$ -- it is the same kind of generalization made for
the potential energy, where only the number of components of the metric
(the scalar product) was increased. By imposing that a quantum must also
satisfy the above relation, it follows immediately that the de Broglie
relation $\lambda = h/p$ is valid in {\it any} $\Re^n$, because Planck's
quantization law did not change. Thus we can also conclude that, as the de
Broglie relation is exact in {\it any} $\Re^n$, the momentum $p$ of the
particle cannot be a function of its coordinate $x$, and so we should
expect that Heisenberg's uncertainty relations are also valid [\cite{Blokhintsev}]. This
result, in a certain sense, properly supports the initial generalization of
the Schr\"odinger equation, a sit should be expected that the equivalence
between Heisenberg's and Schr\"odinger's pictures must be maintained for
other $\Re^n$. This feature is a self-consistency test for this
generalization, which, to our knowledge, has not been used in the past
literature.

          So, it has been shown in this Section that even though the
transition $\Re^3 \rightarrow \Re^n$ does not show ``singular" aspects,
there is a case in which we can still perform it (justified {\it a
posteriori}) and conclude something about the validity of other physical
law in $\Re^n$. The advantage of this procedure was already discussed in
Section 3.

          We will now apply the result of this Section to a particular
physical effect -- the possibility of having thermal neutron diffraction
by crystals in a $\Re^n$ space. It is essentially explained by the de
Broglie hypothesis and then an upper limit for spatial dimensionality will
be obtained, based on the general arguments presented in Section 3.

\section{Thermal neutron diffraction by crystals
as a means for obtaining an upper limit for spatial dimensionality}\label{thermal}

          It is well known that a thermal neutron beam falling onto a
crystal lattice gives rise to diffraction phenomena [\cite{Dunning}], [\cite{Mitchell}] -- known as
``neutron diffraction" -- analogous to those observed when we use incident
$X$-ray  beams. The passage of thermal neutrons through matter gives rise
to scattering processes which are most readily understood in terms of the
wave properties of the neutrons [\cite{Goldberger}], [\cite{Bacon}], [\cite{Wilkinson}]. We define as ``thermal" a neutron
whose kinetic energy corresponds to the mean energy of thermal agitation at
temperature $T$. Usually we can write $p^2/2m \simeq 3K_BT/2$, where the
factor 3 arises when we consider $\Re^3$-space and only 3 degrees of
freedom, corresponding to translational motion, are assumed for the
neutron, {\it i.e.}, by hypothesis, one does not take into account any
internal degree of freedom. Therefore, if we assume the energy
equipartition theorem to be still valid for an $\Re^n$-space, each degree
of freedom will contribute with $K_BT/2$ and the factor 3 should be
replaced by\footnote{Here we are identifying space dimensionality with its number of
translational degrees of freedom.} $n$.

          Since the classical thermodynamics laws do not show singular
aspects concerning the  $\Re^3 \rightarrow \Re^n$ transition\footnote{Assuming time to be one-dimensional (as always assumed in this work)
and ``flowing" in a definite direction. However, the statement made in the
text seems to be no longer true if one tries to develop a thermodynamical
theory in the framework of general relativity. {\it Cf.} Stueckelberg,
E.C.G., ``Thermodynamique dans un continu, riemannien par domaines, et
th\'eor\`eme sur le nombre de dimensions $(d\le 3)$ de l'espace", {\it
Helv. Phys. Acta} {\bf 26} (1953) 417; Stueckelberg, E.C.G. and Wanders,
G., ``Thermodynamique  en Relativit\'e G\'en\'erale", {\it ibid} {\bf 26}
(1953) 307. We thank Dr. M.O. Calv\~ao for pointing out to us these references.} it is
still possible to thermalize a neutron beam in an $\Re^n$-space. Thus the
de Broglie wavelength associanted to the neutron is $\lambda = h/p$, where
$p \simeq (nmk_BT)^{1/2}$. From now on $\lambda$ will be considered as a
function of both the dimensionality of space and neutron velocity
(``temperature"), with $n$ being a parameter to be determined. The starting
point is, therefore, that neutron thermalization may occur in a
$n$-dimensional space. The subsequent process -- neutron diffraction --
simply acts as the {\it detection} of something that has happened inside a
nuclear reactor, for example. To measure $\lambda$ use will be made of
Bragg's law [\cite{Bragg}].

          If $d$ represents the grating spacing the condition for coherent
reflection is given by Bragg's law

$$2d \sin \theta = \ell \lambda, \ \ \ell = 1,2,3, ... $$

          For diffraction patterns to be observed, the wavelength must be
of the order of magnitude of the mean distance between crystallographic
Bragg planes (which in $\Re^3$-space are given by the so called Miller
indexes [\cite{Phillips}], easily generalized to $\Re^n$), but {\it cannot exceed} $2d$.
In this case Bragg's law has no solution for integer $\ell$ and there in no
diffraction pattern.

          The distance $d$ can be measured by using $X$-ray techniques and
thus is independent of the dimensionality of space, {\it i.e.}, for
$X$-rays the relation  $p \simeq (nmk_BT)^{1/2}$, valid for massive
particles (as neutrons, helium atoms, hydrogen molecules {\it etc.}), is
obviously not valid any longer.

          We can then conclude that, in an $\Re^n$-space, diffraction
gratings [\cite{Jenkins}] do exist -- the spacing grating being independent of $n$ --
and it is possible to thermalize a neutron beam. However, it is still
possible to ``measure" $n$ even in the limiting case where a
``one-dimensional crystal" is used as a ``rod" because $\lambda$ is, by
definition, ``one dimensional" and the knowledge of $n$ comes through the
measure os $\lambda$. Thus the above requirement that diffraction gratings
exist in $\Re^n$ seems to be superfluous. In any case, the
3-dimensionality of the macroscopic crystal does not necessarily say
anything about the space dimensionality of the microscopic characteristic
length of thermal neutron production. This information is carried out by
the neutron and will be revealed by the crystal lattice. To make the point,
we are taking into account the possibility that the space dimensionality
may be dependent on the spatial scale (or energy scale) we are probing.

          In its application to solid state problems, neutron diffraction
is similar in theory and experiment to $X$-ray diffraction but, in fact,
regarding some particular aspects, they could be considered as two
complementary techniques [\cite{Goldberger}], [\cite{Bacon}], [\cite{Wilkinson}]. The experimental apparatus we will consider
consists of a monochromatic neutron beam (obtained with usual techniques
[\cite{Bacon}], [\cite{Wilkinson}]) and a crystal. The mean distances between the Bragg's planes are
measured by using $X$-ray techniques. Given a certain crystal one tries to
determine the larger value of these distances which, in general, lies on an
axis of symmetry of the crystal. The neutron beam is then sent on the
crystal in such a way that it will be diffracted by the parallel planes
having as relative distance the aforementioned value. The advantage of
this procedure will be soon understood.

          It is well known from optics that, even when the number of slits
in a diffraction grating is not very large, the intensity of secondary
maxima in the diffraction pattern is much reduced, compared with the
intensity of principal maxima [\cite{Jenkins}]. In the case of neutron diffraction by a
crystal one has a very large number of ``slits" -- the mean intervals
between atoms -- which, clearly, renders difficult the experimental
observation of a high order spectrum. But this does not mean that they
could not be observed in principle. From Bragg's law it follows that to
have a second order spectrum we must have $\lambda \le d$; for a third
order one we need $\lambda \le 2d/3$, and so one. The condition for having
a diffraction pattern with {\it only} the $\ell$-th. order spectrum is,
therefore, $2/(\ell +1) \le \lambda/d \le 2/\ell$. The possible ranges for
the neutron wavelength are then always different and this is an important
point, as we will see now.

          Suppose one can vary (increase) the number of spatial dimensions
for a given constant temperature; for example from $n=3$ to $n=12$. As
$\lambda$ is proportional to $1/\sqrt {n}$, this corresponds to dividing
the wavelength by a factor 2 and, therefore, it is equivalent to going from
a spectrum of order $\ell$ to one of order $2\ell$. This fact, naturally,
strongly suggests that one should observe the {\it first} order spectrum,
as far as one is looking for an upper limit for $n$. We can, thus, perform
a {\it gedanken} experiment where it is possible to prepare a monochromatic
neutron beam satisfying the condition $d\le \lambda \le 2d$, by varying the
neutron velocity and, consequently, $\lambda$, which assures us that no
higher order spectra are presented in the diffraction pattern other than
the first one. Only if one can change $\lambda$ by a factor 2 and still
have the same order diffraction pattern is one sure that it is the first
order spectrum that is observed, because we must remember that we are
taking $n$ as an unknown quantity. After being sure that this is the case,
we can then use the relation $\lambda = h (nmK_BT)^{-1/2}$ for determining
$n$. Therefore, we can conclude that the observation of thermal neutrons
diffraction, under the condition $d\le \lambda \le 2d$, can be used to
measure\footnote{Another recent proposal for measuring the number of dimensions of
space-time, which leads to a fractional dimension, can be found in:
Zeilinger, A. and Svozil, K., ``Measuring the Dimension of Space-Time", {\it
Phys. Rev. Lett.} {\bf 54} (1985) 2553. {\it Cf.} also M\"uller B. and
Sch\"afer A., ``Improved Bounds on the Dimension of Space-Time", {\it
Phys. Rev. Lett.} {\bf 56 (1986) 1215; ``Bounds for the Fractal Dimension
of Space}, {\it preprint n.} UFTP 147/1986 (to be published in {\it J.
Phys. A}. Some consequences of a modification of Newton's and Coulomb's
laws, introduced by assuming a non integer value for the spatial number of
dimensions, are examined in: Jarlskog C. and Yndur\'ain F.J., ``Is the
Number of Spatial Dimensions and Integer?", {\it Europhys. Lett.} {\bf 1}
(1986) 51. There, it is inquired how large can the deviations from the
``standard"  $n=3$ value be. Also the recent work by Grassi A., Sironi G.
and Strini, G., ``Fractal Space-Time and Blackbody Radiation", {\it
Astrophys. Space Sci.} {\bf 124} (1986) 203, is aimed at setting upper
limits to such deviations. It should also be mentioned that in a recent
paper of Gasperini, M., ``Broken Lorentz symmetry and the dimension of
space-time", {\it Phys. Lett.} {\bf B180} (1986) 221 it is shown that the
modification of Newtonian potential -- deviation from the $1/r$
gravitational potential -- following from a deviation of the number of
spatial dimensions from the integer value of $3$, can {\it also} be
obtained in the usual four-dimensional context, provided the $SO(3,1)$
gauge symmetry of gravity is broken. Thus this result gives rise to the
possibility of ambiguous interpretations for small deviations of the
Newtonian gravitational law, but does not affect Coulomb's law.} $n$.

          We shall now analyze the available experimental data. It is known
from $X$-ray measurements that a typical value for $d$ in a crystalline
solid is $d \gtrsim 10^{-10}$~m and the characteristic temperature is $T
\approx 300$~K. For neutron beams, from what has been said above, both
values are independent of space dimensionality. This is the fundamental
fact that allows us to use $d\le \lambda \le 2d$, which gives us the
approximate limit $n \lesssim 5$. For a fixed value of the temperature, one
may ask whether a particular crystal whose $d$ value is such as to test
$n=3$ does exist.

          The wave aspect of the phenomenon discussed in this Section might
lead to supplementary restrictions on the value of $n$.

          In classical physics, diffraction effects can be explained on the
basis of a wave theory by the application of Huygens' construction together
with the principle of interference. In $\Re^{2n}$-space it is well known
that Huygens' principle does not hold [\cite{Hadamard}]. It should also be noted that
Hadamard [\cite{Hadamard23}] has shown that the transmission of wave impulses in a
reverberation-free fashion is possible only in space with an {\it odd}
number of spatial dimensions\footnote{ Further, for the transmission of a wave signal to be free of distortion
it can be shown that $n=1$ and $n=3$ are the only possibilities.} and, in these cases, Huygens' principle
is valid for single differential equations of second order with constant
coefficients. However, Hadamard's conjecture states that this theorem holds
even if the coefficients are not constant [\cite{Courant}]. The Huygens' principle in
then expected to be valid in any $\Re^n$-space where $n$ is odd. Now we
shall assume that the classical results discussed in this paragraph remain
valid when we consider the diffraction of matter by crystals --
traditionally explained by de Broglie's hypothesis within quantum
mechanics.\footnote{Furthermore Bragg's law has been obtained in an alternative way,
without using matter waves and, therefore, independently of Huygens'
construction. Indeed, it has been argued by Bush, R.T., {\it in:} ``The de
Broglie wave derivation for material particle diffraction re-examined: a
rederivation without matter waves", {\it Lett. Nuovo Cimento} {\bf 44}
(1985) 683; ``A theory of particle interference based upon the uncertainty
principle, II. Additional consequences", {\it ibid} {\bf 36} (1983) 241,
that a direct particle interpretation based on Heisenberg's uncertainty
principle can be given to the interference pattern produced by a regular
grating.} This point is far from trivial and is now under
investigation. The difficulty comes from the fact that Hadamard's results
apply to d'Alembert equation, of hyperbolic type, while Scrh\"odinger
equation is parabolic. So, within the above assumption, we can conclude
that thermal neutron diffraction gives an upper limit for spatial
dimensionality which is an odd integer less than or of the order of five.

          We hope the {\it gedanken} experiment performed here, may, in
practice, be carried out in the near future.

\section{Concluding remarks}\label{conclusion}

          In this paper we have discussed the validity of applying the
``stability postulate" to the problem of spatial dimensions. It was shown
that this kind of approach naturally favor {\it a priori} $n=3$. An
alternative approach is proposed where, basically, it is suggested that one
must first demonstrated that the ultimate law used to derive spatial
dimensionality is valid in generic $\Re^n$, rather than simply postulating
the validity of the same equation for an arbitrary $\Re^n$. From this
approach one finds that the constraint obtained on the spatial dimensions
are not only weaker (upper limits) than those obtained by using stability
arguments, but have also a different nature, which we consider more
appropriate to this problem. The main advantage of our methodology is that
it is able to bypass an essential difficulty inherent to the problem to the
problem of the number of spatial dimensions, namely: $n=3$ is never
questioned {\it a priori} when a physical law is established. Clearly it is
not our scope to deduce the number of dimensions of space from a pure
conceptual law [\cite{Grassmann}], but provide a constructive scheme to get at it. As
stated in the Introduction, we believe that the structure of physical space
-- in particular its dimensionality -- is a function of our conceptual
scheme but it does not seem possible to deduce the spatial dimensionality
from it. In the last analysis, one should resort to phenomenology to
determine it.

          In this paper, the fundamental equations generalized to $\Re^n$
were the Schr\"odinger equation and the Einstein energy-mass relation. The
validity of the de Broglie relation for any $\Re^n$ properly supports the
initial generalization of the Schr\"odinger equation (Section~\ref{black}) and, at
the same time, gives a justification for it, in general not found in other
cases. Then, using this result, we have suggested the phenomenon of thermal
neutron diffraction by crystals as a means to determine the number of
spatial dimensions. As a consequence, we have found an upper limit for $n$,
which is an odd integer (by assumption) less than or of the order of five.
We consider the {\it gedanken} experiment performed in Section~\ref{thermal} an {\it
experimentum crucis} for the problem of spatial dimensions and hope it may,
in practice, be carried out in the near future.

          Let us now make some comments about the nature of the different
approaches, considering the physical problem of spatial dimensions, quoted
in this paper. We can divide them into two distinct classes. The first one
corresponds to topological arguments: to this class belong Whitrow's
bio-topological argument [\cite{Whitrow}] and Poincar\'e's argument, based on the {\it
analysis situs} [\cite{Poincare17}]. The kind of constraints obtained from it is a
lower limit for spatial dimensionality, {\it e.g.}, $n\ge 3$. In the second
class, we group all other arguments where it is necessary to introduce a
metric space and this seems to restrict the range of possible values of
$n$. A metric space is introduced whenever we consider the existence of an
interacting system as the starting-point in the discussion of the problem
of spatial dimensions. It is clear to begin with an interacting system,
knowledge of the form of the interaction -- the physical law describing
the phenomenon in a space-time manifold -- is a necessary condition. This
renders the class of ``metric arguments" more ``complete" {\it a priori},
in the sense that it contains more information than the class of ``purely
topological arguments". The difference can be considered as the cause of
the difference between the two types (or classes) of constraints for $n$.
There is, however, an exception to this general picture that should be
emphasized: the Maxwell electromagnetic theory. We would like to point out
here its distinguished r\^ole in the physical problem of space dimensions.
All the attempts to obtain the space dimensionality which are based upon
the structure of Maxwell's equations (no matter whether they belong to the
class of metric approach or not) give $n=3$.

          It is not perhaps out of place to present now some almost obvious
remarks about time (and space) ``scale" of the arguments previously
discussed. Ehrenfest's stability argument is valid for distances of the
order of the solar system and in a time scale large enough to make the
evolution of life possible on Earth (as mentioned by Whitrow\footnote{One may add the following remark to Whitrow's argument about this
subject [\cite{Whitrow}]. It is not sufficient that the intensity of solar radiation
on Earth's surface should not have fluctuated greatly for still having life
on Earth. The fact that Sun's spectra of radiation did not fluctuate very
much is also required.}).
Tangherlini's work about the stability of $H$ atoms can be invoked here to
suggest the validity of chemistry in the same time scale as a necessary,
although not sufficient, condition -- at least chemical thermodynamics of
irreversible process should be also valid. The presence of atomic spectra
in remote stars may also indicate that space has had the same
dimensionality at cosmic scale. To have such a cosmic constraint on space
dimensionality is very interesting and we hope to treat this point in a
future communication.

          It is also interesting to note that all the arguments presented
up to now that depend ion the presence of matter are essentially metric.
This is the case of Ehrenfest-Tangherlini-Whitrow. Topological arguments
are basically related to the idea of a field -- this is the case of
Maxwell's theory, as mentioned before, and Wien's law, which involves,
essentially, the equilibrium of radiation.

          As for most physical arguments used to obtain the spatial
dimensionality it is necessary to introduce a metric space, our two last
critical comments are dedicated to clarify some aspects involving it.

          Firstly, in [\cite{Tangherlini63}] the author was led to conclude that the
stability postulate, applied to the N.K. problem, fixes the dimensionality
of space and, at the same time, is an absolute prerequisite for a
comparison of relative distances between bodies to be physically possible.
However, taking into account the analysis we have done and the example we
have proposed in the preceding Sections. we are led to conclude that, in
fact, it is the physical interaction between two bodies, or two systems,
that necessarily leads to the introduction of a metric-space in order to
be able to obtain the number of spatial dimensions in these cases; but
neither the stability postulate nor a metric-space [\cite{Schonberg}] are indeed {\it
necessary} to fix the dimensionality of space.

          Secondly, we would like to point out that the necessity to have a
metric-space for most physical arguments concerning the problem of spatial
dimensions, brings with itself the notion of distance, traditionally based
on the differential homogeneous quadratic form, $ds^2 = g_{\mu\nu} dx^\mu
dx^\nu$, which, in the last analysis, is an arbitrary choice -- indeed
there is no logical argument for excluding other forms for the line element
as $ds^4$, $ds^6$, $ds^8$ {\it etc}. In spite of this (up to now) logical
impossibility the importance of investigating the nature of the exponent 2
was emphasized in an early work by Ehrenfest [1920]. His conjecture that
this 2~could be related to the dimensionality of space is, however, yet to
be demonstrated. Nevertheless, so far as the formula for a line element
in a manifold of $n$ dimensions is viewed as arbitrary, some care must
clearly be exercised in advancing Ehrenfest's conjecture. If, on the
contrary, this conjecture is shown to be actually true, we can ask whether
it can be related in some way to Fermat's last theorem.

          In conclusion, we would like to say that although some
epistemological difficulties concerning the use of ``stability arguments"
are bypassed by the methodology proposed in this paper, there remains,
somehow, a certain incompleteness since a physical event takes place not
only in space, but in {\it space-time}. Thus the problem of the number of
space dimensions and that of time dimensions are probably not independent.
One can then ask whether it is possible to propose a more general
methodology which could be able to constrain not only the number of spatial
dimensions but also, simultaneously, time dimensionality. Are these numbers
actually related? Is it possible to prove time to be one-dimensional by
disclosing space dimensionality and/or vice-versa? It is our conviction
that, in the future, further efforts should be made trying to answer these
questions, whether or not a deeper comprehension on the problem of space
dimensionality is to be reached.
\vskip 0.5 true cm

\noindent {\it Acknowledgments:} One of us (F.C.) would like to thank Prof.
Enrico Predazzi for his kind hospitality at the Dipartimento di Fisica
Teorica dell'Univerisit\`a do Torino where this work was completed. We are
also grateful to him -- as well as to Dr. Leonardo Castellani -- for a
critical reading of the manuscript, for lively discussions and useful
suggestions. We wish to thank Dr. Eduardo Valadares for the translation of
Ehrenfest's German paper and fruitful discussions, and Dr. Roberto Nicolsky for useful
comments concerning Sec.~\ref{thermal}. The authors are indebted to Mrs.~Regina Moura,
Profs.~Jo\~ao dos Anjos and Erasmo Ferreira for providing them with copies
of several references. F.C. thanks the Conselho Nacional de Desenvolvimento
Cient\'{\i}fico e Tecnol\'ogico (CNPq) of Brazil for financial support.

\noindent {\bf R\'esum\'e} --
Pourquoi l'espace a-t-il trois dimensions? La premi\`ere r\'eponse \`a
cette question, compl\`etement fond\'ee sur des raisons physiques, fut
don\'ee par Ehrenfest en 1917, qui montra que la condition de stabilit\'e
pour un syst\`eme plan\'etaire \`a deux corps \`a $n$-dimensions pose des
contraintes tr\`es puissantes sur la dimensionalit\'e de l'espace et
favorise 3-d. Cette approche du probl\`eme sera d\'enomm\'ee ``postulat de
stabilit\'e" dans cet article et, comme le montra Tangherlini en 1963, elle
est ancore valable dans le domaine de la relativit\'e g\'en\'erale ausi
bien que pour l'atome d'hydrog\`ene quantique, en donnant toujours la
m\^eme contrainte pour la dimensionalit\'e de l'espace. Dans ce travail,
avant de faire une analyse critique de la m\'ethodologie rappel\'ee
ci-dessus, nous faisons une br\`eve discussion pour souligner et clarifier
quelques aspects physique g\'en\'eraux du probl\`eme relatif \`a la
determination de la dimensionalit\'e de l'espace. Ensuite, les
cons\'equences \'epist\'emologiques de la m\'ethodologie d'Ehrenfest seront
revues de fa\c con critique. On propose un proc\'ed\'e alternatif pour
arriver \`a d\'eterminer le nombre de dimensions correct, dans lequel le
postulat de stabilit\'e (et les singularit\'es implicites dans la physique
\`a trois dimensions) ne constitue pas une partie essentielle de
l'argumentation. De cette mani\`ere, les principaux probl\`emes
\'epist\'emologiques contenus dans l'id\'ee originale d'Ehrenfest sont
\'evit\'es. La m\'ethodologie alternative propos\'ee dans ce travail est
b\^atie sur la r\'ealisation et la discussion de la th\'eorie quantique \`a
$n$-dimensions exprim\'ee par la loi de Planck, la formule de de Broglie
et la relation d'incertitude de Heisenberg. Par cons\'equent, il est
possible de proposer une exp\'erience, bas\'ee sur la diffraction des
neutrons thermiques par des cristaux, pour mesurer directement la
dimensionalit\'e de l'espace. Finalement, le r\^ole particulier jou\'e par
la th\'eorie \'electromagn\'etique de Maxwell pour la d\'etermination de la
dim\'ensionalit\'e de l'espace est soulign\'e.

%%%%%%%%%%%%%%%%%%%%%%%%%%%%%%%%%%%%%%%%%%%%%%%%%%%%%%%%%%%%%%%%%%%%%%%%%

%\clearpage


\begin{thebibliography}{}

\bibitem[Jammer, M.\hspace*{-.2cm} (1954)]{Jammer} Jammer, M., 1954 - For a brief historical discussion of these works see the book
{\it Concepts of Space: the History of Theories of Space in Physics}. Cambridge, Mass.: Harvard Univ. Press,  Chapter~5.

\bibitem[Kant, I.\hspace*{-.2cm} (1747)]{Handyside} Kant, I., 1747 - {\it Gedanken von der wahren Sch\"atzung der lebendigen Kr\"afte und Beurtheilung der Beweise, deren sich Herr von Leibniz und andere Mechaniker in dieser Streitsache bedient haben, nebst einigen vorhergehenden Betrachtungen, welche die Kraft der K\"orper \"uberhaupt betreffen}, K\"onigs\-berg, 1747; reprinted in: Kant {\it Werke}, Band 1, {\it Vorkritische Schriften}, Wissenschaftliche Buchgesellschaft Darmstadt, 1983. English translation of part of this work was done by J. Handyside and published as part of the book {\it Kant's inaugural dissertation and the early writings on space}. Chicago: Open Court, 1929, reprinted by Hyperion Press, 1979,  pp.~11-15.

\bibitem[Poincar\'e, H.\hspace*{-.2cm} (1913)]{Poincare} Poincar\'e, H., 1913 - {\it La valeur de la Science}, Paris,
Flammarion, pp.~96-136.

\bibitem[B\"uchel, W.\hspace*{-.2cm} (1963)]{Buchel} B\"uchel W., 1963 - ``Warun hat der Raum drei
Dimensionen?", {\it Physikalische Bl\"atter} {\bf 19}, No.~2, pp.~547-49, translated and adapted by Freeman I.M., ``Why is Space Three-Dimensional?", {\it American Journal of Physics} {\bf 37} (1969) pp.~1222-1224.

\bibitem[Gr\"unbaum, A.\hspace*{-.2cm} (1974)]{Grunbaum} Gr\"unbaum, A., 1974 - {\it Philosophical Problems of Space and Time}, second,
enlarged edition. Dordrecht: D. Reidel Publ. Co., pp.~330-37, pp.~833-34, and refs. therein.

\bibitem[Barrow, J.D.\hspace*{-.2cm} (1983)]{Barrow}  Barrow, J.D., 1983 - ``Dimensionality", {\it Phil.
Trans. Roy. Soc. London} {\bf A310}, pp.~337-346; {\it Cf.} also Barrow J.D. and Tipler F., {\it The Anthropic Cosmological Principle}, Oxford, Claredon Press, 1986, Chapter IV.

\bibitem[Jammer, M.\hspace*{-.2cm} (1954)]{Jammer2} Jammer, M., 1954 - {\it op. cit.}, p.~171.

\bibitem[Grassmann, H.\hspace*{-.2cm} (1878)]{Grassmann} Grassmann, H., 1878 - {\it Die Ausdehnungslehre}, 2nd.~ed., Leipzig, p.~XXIII, as quoted by Jammer M., {\it op.~cit.} p.~184.

\bibitem[Ehrenfest, P.\hspace*{-.2cm} (1917)]{Ehrenfest17} Ehrenfest, P., 1917 - ``In what way does it become manifest in the
fundamental laws of physics that space has three dimensions?", {\it Royal Netherlands Academy of Arts and Sciences (KNAW)} {\bf 20}, pp.~200-209; reprinted in  Klein, M.J. (ed.), {\it Paul Ehrenfest - Collected Scientific Papers}. Amsterdam: North Holland Publ. Co., 1959, pp.~400-409.

\bibitem[Ehrenfest, P.\hspace*{-.2cm} (1920)]{Ehrenfest20} Ehrenfest, P., 1920 - ``Welche Rolle spielt die Dreidimensionalit\"at des Raumes in den Grund\-gesetzen der Physik?", {\it Annalen der Physik} {\bf 61}, pp.~440-446.

\bibitem[Whitrow, G.J.\hspace*{-.2cm} (1955)]{Whitrow} Whitrow G.J., 1955 - ``Why Physical Space has
Three Dimensions?", {\it Brit. J. Phil. Sci.} {\bf 6}, pp.~13-31. See also {\it The Structure and Evolution of the
Universe}, New York, Harper and Row, 1959.

\bibitem[Tangherlini, F.R.\hspace*{-.2cm} (1963)]{Tangherlini63} Tangherlini F.R., 1963 - ``Schwarzschild Field in $n$ Dimensions
and the Dimensionality of Space Problem", {\it Nuovo Cimento} {\bf 27}, pp.~636-651.

\bibitem[Tangherlini, F.R.\hspace*{-.2cm} (1986)]{Tangherlini86} Tangherlini F.R., 1986 - ``Dimensionality of
Space and the Pulsating Universe", {\it ibid} {\bf 91B}, pp.~209-217.

\bibitem[Weyl, H.\hspace*{-.2cm} (1918)]{Weyl} Weyl H., 1918 - ``Gravitation und Elektrizit\"at", {\it
Sitzungsberichte d. Preuss. Akad. d. Wiss. Berlin}, \textbf{26}, pp.~465-480; ``Eine neue
Erweiterung der Relativit\"atstheorie", {\it Ann. Physik} {\bf 59} (1919)
pp.~101-133; See also his {\it Space, Time, Matter}, transl. by H.L. Brose, New York, Dover,
1952, pp.~282-285.

\bibitem[Bergmann, P.G.\hspace*{-.2cm} (1942)]{Bergmann} Bergmann P.G., 1942 - {\it Introduction to the Theory of Relativity},
New York, Prentice-Halll; ``Unitary Field Theory", {\it Physics Today}, march 1979, pp.~44-51.

\bibitem[Dirac, P.A.M.\hspace*{-.2cm} (1973)]{Dirac} Dirac P.A.M., 1973 - ``Fundamental Constants and their Development in Time", {\it in} J. Mehra (ed.), {\it The
Physicist's Conception of Nature}, Dordrecht, D. Reidel Publ. Co., pp.~52-53, where it is argued that there is no objection to Weyl's theory.

\bibitem[Kaluza, Th.\hspace*{-.2cm} (1921)]{Kaluza} Kaluza, Th., 1921 - ``Zum Unit\"atsproblem der Physik", {\it
Sitzungsber. d. Preuss. Akad. d. Wiss., Berlin, Math. Physik} {\bf 1}, pp.~966-972.

\bibitem[Klein, O.\hspace*{-.2cm} (1926)]{Klein} Klein O., 1926 - ``Quantum Theory and the Five-Dimensional Relativity
Theory", {\it Zeit. f\"ur Physik} {\bf 37}, pp.~895-906.

\bibitem[Penney, R.\hspace*{-.2cm} (1965)]{Penney}  Penney R., 1965 - ``On the Dimensionality of the Real World", {\it
Journ. Math. Phys.} {\bf 6}, pp.~1607-1611.

\bibitem[Sch\"onberg, M.\hspace*{-.2cm} (1971)]{Schonberg} Sch\"onberg M., 1971 - ``Electromagnetism and Gravitation", {\it Revista
Brasileira de F\'{\i}sica} {\bf 1}, p.~91 and refs. therein. See also Sch\"onberg's book {\it Pensando a F\'{\i}sica}. S\~ao Paulo: Brasiliense,
1985, pp.~89-91.

\bibitem[Van Niewenhuysen, P.\hspace*{-.2cm} (1981)]{Niewenhuysen} See, for example, the review article of Van Niewenhuysen
P., 1985 - ``Supergravity", {\it Phys. Rep.} {\bf C68} (4), pp.~189-398. We thank Dr. M. Gasperini for pointing out to us this reference.

\bibitem[Schwarz, J.H.\hspace*{-.2cm} (1985)] {Schwarz} Schwarz, J.H., 1985 -
``Introduction to superstrings", {\it in} {\it Superstrings and Supergravity -- Proceedings of the Twenty-Eighth Scottish Univ. Summer
School in Phys.}, A.T. Davies and D.G. Sutherland (Eds.), Oxford, Univ. Printing House, 1985, pp.~95-124.

\bibitem[Mansouri, F. and Witten, L.\hspace*{-.2cm} (1985)]{Mansouri} Mansouri F. and Witten L., 1985 - ``Can isometries tell us about the
extra dimensions", {\it in Symposium on Anomalies, Geometry, Topology}, W.A. Bardeen and A.R. White (eds.), Singapore, World Scientific Publ. Co., pp.~509-10.

\bibitem[Poincar\'e, H.\hspace*{-.2cm} (1917)]{Poincare17} Poincar\'e H., 1917 - {\it Derni\`eres Pens\'ees}. Paris: Flammarion. See also [\cite{Poincare}].

\bibitem [Stern, O. and Gerlach, W.\hspace*{-.2cm} (1922)]{Stern} Stern, O. and Gerlach, W., 1922 - ``Experimental Determination of the
Magnetic Moment of Silver Atom", {\it Zeitschrift f\"ur Physik} {\bf 8}, pp.~110-111; ``Experimental Test of the Applicability of the Quantum Theory to the
Magnetic Field", {\it idem} {\bf 9}, pp.~349-352, and ``Magnetic Moment of Silver Atom" {\it id.} {\bf 9}, pp.~353-355.

\bibitem [Blokhintsev, D.\hspace*{-.2cm} (1981)]{Blokhintsev} Blokhintsev, D.,  1981 - {\it Principes de M\'ecanique Quantique},
Moscow, Mir Publ.; {\it Quantum Mechanics}. Dordrecht: Reidel, 1964, par.~15, pp.~44-45, translated from the third and fourth Russian editions
by J.B. Sykes and M.J. Kearsley.

\bibitem [Dunning, J.R. {\it et al.}\hspace*{-.2cm} (1935)]{Dunning} See, for example, Dunning, J.R.; Pegram, G.B.; Fink, G.A.;
Whitehall, D.P. and Segr\`e, E., 1935 - ``Velocity of Slow Neutrons by Mechanical Velocity Selector", {\it Phys. Rev.} {\bf 48}, p.~704.

\bibitem [Mitchell, D.P. and Powers, P.N.\hspace*{-.2cm} (1936)]{Mitchell} Mitchell, D.P. and
Powers, P.N., 1936 - ``Bragg Reflection of Slow Neutrons", {\it Phys. Rev.} {\bf 50}, pp.~486-487.

\bibitem [Goldberger, M. and Seitz, F.\hspace*{-.2cm} (1936)]{Goldberger} Goldberger M. and Seitz F., 1947 - ``Theory of the Refraction and
the Diffraction of Neutrons by Crystals", {\it Phys. Rev.} {\bf 71}, pp.~294-310.

\bibitem [Bacon, G.E.\hspace*{-.2cm} (1962)]{Bacon} Bacon G.E., 1962 - {\it Neutron Diffraction}. New York: Oxford Univ.
Press.

\bibitem [Wilkinson, M.K., Wollan, E.O. and Koehler, W.C.\hspace*{-.2cm} (1961)]{Wilkinson} Wilkinson, M.K.; Wollan, E.O. and Koehler, W.C., 1961 - ``Neutron
Diffraction", {\it Ann. Rev. Nucl. Sci.} {\bf 11}, pp.~303-348 and refs.
therein.

\bibitem [Bragg, W.L.\hspace*{-.2cm} (1913)]{Bragg} Bragg, W.L., 1913 - ``The Diffraction of Short Electromagnetic Waves
by a Crystal", {\it Proc. Cambridge Phil. Soc.} {\bf 17}, pp.~43-57; Bragg, W.H. and Bragg, W.L., ``The Reflection of $X$-rays by Crystals", {\it Proc.
Royal Soc. London} {\bf A88}, pp.~428-438.

\bibitem [Phillips, F.C.\hspace*{-.2cm} (1946)]{Phillips} Phillips, F.C., 1946 - ``An Introduction to Crystallography". London: Longman.

\bibitem [Jenkins, F.A. and White, H.E.\hspace*{-.2cm} (1937)]{Jenkins} Jenkins, F.A. and White, H.E., 1937 - {\it Fundamentals of Physical
Optics}, first ed., twelfth impression. New York: McGraw-Hill, 1937, chap.~7, p.~146.

\bibitem [Hadamard, J.\hspace*{-.2cm} (1903)]{Hadamard} See, for example, Hadamard, J., 1903 - {\it Le\c cons sur la
propagation des Ondes}. Paris: Hermann, chap. VII, par.~3; ``Les
probl\`emes aux limites dans la th\'eorie des \'equations aux d\'eriv\'ees
partielles", {\it Bull. Soc. Franc. de Phys.}, 1906, pp.~276-315, and
references therein; ``Th\'eories des \'equations aux d\'eriv\'ees partielles lin\'eaires hyperbolique et du probl\`eme de Cauchy", {\it Acta
Math.} {\bf 31} (1908) 333.

\bibitem [Hadamard, J.\hspace*{-.2cm} (1923)]{Hadamard23} Hadamard, J., 1923 - {\it Lectures on Cauchy's problem in linear
partial differential equations}. New Haven: Yale Univ. Press, 1923, pp.~53-54, 175-77 and 235-36.

\bibitem [Courant, R. and Hilbert, D.\hspace*{-.2cm} (1962)]{Courant} Courant, R. and Hilbert, D., 1962 - {\it Methods of Mathematical
Physics}, vol.~II. New York: Interscience Publ., p.~765 and notes therein.

\end{thebibliography}
\end{document}